\documentclass[12pt]{article}
\usepackage{stmaryrd}
\usepackage{amscd}
\usepackage{fancyhdr}
\usepackage{titlesec}
\usepackage{amssymb}
\usepackage{ifthen}
\usepackage{appendix}
\usepackage{bm}
\usepackage{setspace}

\usepackage{amsmath}
\usepackage{amssymb}
\usepackage{latexsym,floatrow}
\usepackage{bbm}
\usepackage{verbatim}
\usepackage{amsfonts}
\usepackage{epsfig}
\usepackage{tikz}
\usepackage{bbm} 
\usepackage{amssymb,amsmath,tikz,graphicx,float,xspace,chemarrow}
\usepackage{graphicx,floatrow}
\usepackage[utf8]{inputenc}
\makeatletter

\newcommand{\be}{\begin{equation}}
\newcommand{\ee}{\end{equation}}
\newcommand{\bea}{\begin{eqnarray}}
\newcommand{\eea}{\end{eqnarray}}

\newcommand{\distas}[1]{\mathbin{\overset{#1}{\kern\z@\sim}}}%
\newsavebox{\mybox}\newsavebox{\mysim}
\newcommand{\distras}[1]{%
  \savebox{\mybox}{\hbox{\kern3pt$\scriptstyle#1$\kern3pt}}%
  \savebox{\mysim}{\hbox{$\sim$}}%
  \mathbin{\overset{#1}{\kern\z@\resizebox{\wd\mybox}{\ht\mysim}{$\sim$}}}%
}

\newcommand{\y}{\mathbf{y}}

\newcommand{\btheta}{\mbox{\boldmath $\theta$}}
\newcommand{\hbtheta}{\mbox{\boldmath ${\hat \theta}$}}

\begin{document}
\begin{center} { \Large \sc A Class of Skewed Distributions with Applications in Environmental Data}
\end{center}

\begin{center}
{\large   Indranil Ghosh$^{1}$,\quad  Hon Keung Tony Ng$^{2}$ }
\\

\vspace*{0.5cm}
{\large $^{1}$ University of North Carolina, Wilmington, North Carolina, USA \\
$^{2}$ Southern Methodist University, Dallas, Texas, USA.}

\end{center}

\vspace*{0.5cm}

\begin{abstract}
In environmental studies, many data are typically skewed and it is desired to have a flexible statistical model for this kind of data. In this paper, we study a class of skewed distributions by invoking arguments as described by Ferreira and Steel (2006, {\it Journal of the American Statistical Association}, 101: 823--829). In particular, we consider using the logistic kernel to derive a class of univariate distribution called the truncated-logistic skew symmetric (TLSS) distribution. We provide some structural properties of the proposed distribution and develop the statistical inference for the TLSS distribution. A simulation study is conducted to investigate the efficacy of the maximum likelihood method. For illustrative purposes, two real data sets from environmental studies are used to exhibit the applicability of such a model.
\end{abstract}



\section{Introduction}
\label{S:1}

The need for skewed distributions arises in every area of the sciences, engineering and medicine because data are likely coming from asymmetrical populations.
One of the common approaches for the construction of skewed distributions is to introduce skewness into some known symmetric distributions. Ferreira and Steel (2006) presented a unified approach for constructing such a class of skewed distributions. Let $X$ be a symmetric random variable about zero with probability density function (pdf) $f_{X}(\cdot)$ and cumulative distribution function (cdf) $F_{X}(\cdot)$. Then, the random variable $Y$ is a skewed version of the symmetric random variable $X$ with pdf
\begin{equation}
f_{Y}(y) = f_{X}(y) w\left[ F_{X}(y) \right],  \quad y \in \mathbb{R},
\end{equation}
where $w[\cdot]$ is a pdf defined on the unit interval $(0, 1)$ (Definition 1, Ferreira and Steel, 2006).
The unified family of distributions defined in Eq. (1.1) contains many well-known families of skewed distributions. One of the commonly used class of skewed distributions in the form of Eq. (1.1) is the skewed distributions introduced by Azzalini (1985). Specifically, take $w(x) = 2F_{X} \left[ \lambda F^{-1}_{X}(x)\right].$ Then, Eq. (1.1) reduces to
\begin{equation}
f_{Y}(y) = 2 f_{X}(y) F_{X}(\lambda y),  \quad y \in \mathbb{R}, \quad \lambda \in \mathbb{R}.
\end{equation}
A particular case of the model in Eq. (1.2) is the skewed normal distribution obtained by setting $f_{X}(\cdot) = \phi(\cdot)$ and $F_{X} (\cdot) = \Phi(\cdot)$, where $\phi(\cdot)$ and $\Phi(\cdot)$ are the pdf and cdf of the standard normal distribution, respectively. The family of distributions given by Eq. (1.2) and the skew-normal class have been studied and extended by many authors, for example, see Azzalini (1986), Azzalini and Dalla Valle (1996), Azzalini and Capitanio (1999), Arnold and Beaver (2000), Pewsey (2000), Loperfido (2001), Arnold and Beaver (2002), Nadarajah and Kotz (2003), Gupta and Gupta (2004), Behboodian et al. (2006), Nadarajah and Kotz (2006), Huang and Chen (2007) and Sharafi and Behboodian (2006).

In environmental studies, data are typically skewed and different skewed distributions such as the Weibull, lognormal and gamma distributions are often used to model such data sets (see, for example, EPA 1992, Singh et al. 2002). For instance, the soil concentrations of the contaminants of potential concern (Singh et al. 2002; Shoari et al. 2015), mercury concentration in swordfish (Lee and Krutchkoff 1980), and survival time times of mice exposed to gamma radiation (Gross and Clark 1975; Grice and Bain 1980) are fitted by different skewed distributions.
In this paper, we aim to propose a class of skew-symmetric distributions as an alternative model for fitting skewed data originating from various environmental applications.

This paper is organized as follows. In Section 2, we discuss the proposed class of skew symmetric distributions and introduce some special cases of this class of distributions. In Section 3, we study the structural properties of the proposed class of distributions.
Then, the random number generation of the proposed class of distributions is discussed in Section 4.
In Section 5, the maximum likelihood estimation method is used to estimation the model parameters of the proposed class of distributions.
In Section 6, two real data sets from environmental studies are used to illustrate the usefulness of the proposed class of distributions. Finally, some concluding remarks are presented in Section 7.

\section{Truncated Logistic Skew-Symmetric Family of Distributions}
\setcounter{equation}{0}

In this section, we introduce the truncated logistic skew-symmetric (TLSS) family of distributions and study various structural properties of this family. At first, we provide the definition of the proposed family of distributions as follows.

\noindent
{\bf Definition 1.} A random variable  $Y$ has the truncated-logistic skew-symmetric distribution with parameter $\lambda,$ namely,  $TLSS(\lambda),$ if its pdf has the following form:
\begin{equation}
f_Y(y; \lambda) = \bigg[\frac{2(1+e^{-\lambda})}{(1-e^{-\lambda})}\bigg] \bigg[ \frac{\lambda f_X(y) e^{-\lambda F_X(y)}}{(1+e^{-\lambda F_X(y)})^2} \bigg], \qquad y\in \mathbb{R}, \quad \lambda \in \mathbb{R},
\label{pdf}
\end{equation}
where $f_X(\cdot)$ and $F_X(\cdot)$ are, respectively, the pdf and the cdf of a symmetric random variable $X$ about zero, and $\lambda$  is a shape parameter.

\noindent From Eq. (\ref{pdf}), the associated cdf of the random variable $Y$ has the form
\begin{equation}
F_Y(y;\lambda) = \int^{y} _{-\infty} f_Y(t;\lambda) dt = \bigg[ \frac{1-e^{-\lambda F_X(y)}}{1+e^{-\lambda F_X(y)}}\bigg] \bigg(\frac{1+e^{-\lambda}}{1-e^{-\lambda}} \bigg), \qquad y \in \mathbb{R}, \quad \lambda \in \mathbb{R}.
\label{cdf}
\end{equation}
Then, from Eq. (\ref{cdf}), the inverse cdf of $Y$ can be expressed as
\begin{equation}
F^{-1}_Y(u;\lambda)
= F^{-1}_X \bigg( \frac{1}{\lambda}\log\left[ \frac{1-u(\frac{1+e^{-\lambda}}{1-e^{-\lambda}})}{u(\frac{1+e^{-\lambda}}{1-e^{-\lambda}})}\right] \bigg), \qquad u \in (0, 1), \quad \lambda \in \mathbb{R}.
\label{icdf}
\end{equation}
The inverse cdf in Eq. (\ref{icdf}) can be used to obtain the distribution quantiles. Specifically, if $\Pr\left(Y \leq\xi_{p}\right) = p$,  for any $p\in (0,1]$, then the $p$-th quantile, $\xi_{p},$ can be obtained by using Eq. (\ref{icdf}).
In addition, the inverse cdf in Eq. (\ref{icdf}) can be used to generate random sample from $TLSS(\lambda)$ based on a uniform random number in $(0, 1)$ by means of the inverse transform method, i.e., $Y = F_{Y}^{-1}(U; \lambda)$, where $U$ is a random number from uniform distribution in (0, 1) (see Section 4 for the details). 

Note that the class of distributions defined in Eq. (\ref{pdf}) is a particular case of the class in Eq. (1.1) with $$w(x) = \bigg[\frac{2(1+e^{-\lambda})}{(1-e^{-\lambda})}\bigg] \frac{e^{-\lambda x}} {\left(1+e^{-\lambda x}\right)^{2}},$$
which is the pdf of a truncated logistic distribution. By introducing the logistic function and replacing $F_{X}(\lambda y)$ by $\lambda F_{X}(y),$ one can see that the family of distributions in Eq. (\ref{pdf}) is a natural extension of Eq. (1.1) to a logistic family. Furthermore, the family of distributions in Eq. (\ref{pdf}) is symmetric with respect to $\lambda$ in the sense that  $f (y; \lambda) = f (-y; -\lambda).$ Additionally, in the limit, as  $\lambda\rightarrow 0,$  $Y \sim TLSS(\lambda) $ has the same distribution as $X.$ Again, we remark that Eq. (\ref{pdf}) is undefined at  $\lambda = 0,$  so $\lambda = 0$ should be interpreted as the limit  $\lambda \rightarrow 0.$ If  $\lambda \rightarrow (\pm)\infty,$ then $Y \sim TLSS(\lambda)$ reduces to  degenerate random variables. If  $\lambda\rightarrow \infty$,  then  $F_{Y}(y)=0$  if $F_{X}(y)=0$ and  $F_{Y}(y)=1 $ for all other values of $y.$ If $\lambda \rightarrow -\infty$ then $F_{Y} (y) = 1$ if  $F_{X}(y) = 1$ and $F_{Y} (y) = 0$ for all other values of $y.$

Next, we consider some specific members of the TLSS family:
\begin{enumerate}
\item If $X \sim {\mbox {Normal}}(\mu, \sigma)$, i.e., $f_{X}(x)=\phi(\frac{x-\mu}{\sigma})$ and $F_{X}(x)=\Phi(\frac{x-\mu}{\sigma})$, $x \in \mathbb{R}$, $\mu \in \mathbb{R}$ and $\sigma \in \mathbb{R}^{+}$, in Definition 1, then Eq. (2.3) gives the pdf of the random variable $Y$ as
\begin{eqnarray}
f_Y(y; \mu, \sigma, \lambda)= \bigg[\frac{2(1+e^{-\lambda})}{\sigma(1-e^{-\lambda})}\bigg] \bigg\{ \frac{\lambda \phi(\frac{y-\mu}{\sigma})  e^{-\lambda \Phi(\frac{y-\mu}{\sigma})}}{[1+e^{-\lambda \Phi(\frac{y-\mu}{\sigma})}]^2} \bigg\}, y \in \mathbb{R},  \lambda \in \mathbb{R}, \mu \in \mathbb{R}, \sigma\in \mathbb{R^{+}}.
\label{tlsn}
\end{eqnarray}
We refer the distribution in Eq. (\ref{tlsn}) as the truncated-logistic-skew normal (TLSN) distribution with parameters $\lambda, \mu$ and $\sigma$.
In Figure 1, we plotted the pdfs of the TLSN distribution with $\mu = 0$ and $\sigma = 1$ for different values of the parameter $\lambda$.

\item  If $X \sim {\mbox {Laplace}}(\mu, b)$, i.e., $f_{X}(x)=\frac{1}{2b}\exp\left(-\frac{x-\mu}{b}\right)$ and \linebreak
$F_{X}(x)=\frac{1}{2}+\frac{1}{2}sgn(x-\mu)\left[1-\exp\left(-\frac{x-\mu}{b}\right)\right]$, $x \in \mathbb{R}$, $\mu \in \mathbb{R}$ and $b \in \mathbb{R}^{+}$, in Definition 1, then Eq. (\ref{pdf}) gives the pdf of the random variable $Y$ as
\begin{eqnarray}
\nonumber
f_Y(y; \mu, b, \lambda) & = & \bigg[\frac{2(1+e^{-\lambda})}{(1-e^{-\lambda})} \bigg] \bigg\{ \frac{\lambda \left(\frac{1}{2b}\exp\left(-\frac{x-\mu}{b}\right) \right) e^{-\lambda\left(\frac{1}{2}+\frac{1}{2}sgn(x-\mu)\left[1-\exp\left(-\frac{x-\mu}{b}\right)\right] \right)}}{[1+e^{-\lambda \left( \frac{1}{2}+\frac{1}{2}sgn(x-\mu)\left[1-\exp\left(-\frac{x-\mu}{b}\right)\right]\right)}]^2} \bigg\},\\
& & \qquad \qquad \qquad \qquad \qquad \qquad \qquad y \in \mathbb{R}, \lambda\in \mathbb{R}, \mu \in \mathbb{R}, b \in \mathbb{R^{+}}.
\label{tlsl}
\end{eqnarray}
We refer the distribution in Eq. (\ref{tlsl}) as the truncated-logistic-skew Laplace (TLSL) distribution with parameters  $\lambda, \mu$ and $b$.
In Figure 2, we plotted the pdfs of the TLSL distribution with $\mu = 0$ and $b = 1$ for different values of the parameter $\lambda$.

\item If $X \sim \text{Cauchy}(\mu, \xi)$, i.e., $f_{X}(x)=\frac{1}{\pi \left[1+\left(\frac{x-\mu}{\xi}\right)^{2}\right]}$ and
$F_{X}(x)=\frac{1}{2}+\frac{1}{\pi}\arctan\left(\frac{x-\mu}{\xi}\right)$,
$x \in \mathbb{R}$, $\mu \in \mathbb{R}$ and $\xi \in \mathbb{R}^{+}$, in Definition 1, then Eq. (2.3) gives the pdf of the random variable $Y$ as
\begin{eqnarray}
\nonumber
f_Y(y; \mu, \xi, \lambda) & = & \bigg[\frac{2(1+e^{-\lambda})}{(1-e^{-\lambda})}\bigg]\\
\nonumber
& & \times \Bigg[\frac{\lambda \exp\left(-\lambda \left[\frac{1}{2}+\frac{1}{\pi}\arctan\left(\frac{x-\mu}{\xi}\right)\right]\right)}
{\left\{\pi\xi \left[1+\left(\frac{x-\mu}{\xi}\right)^{2}\right]\right\}\left(1+ \left[1+\exp\left(-\lambda \left[\frac{1}{2}+\frac{1}{\pi}\arctan\left(\frac{x-\mu}{\xi}\right)\right]\right)\right]\right)^{2}}\Bigg],\\
& & \qquad \qquad \qquad \qquad \qquad \qquad \qquad y \in \mathbb{R}, \lambda\in \mathbb{R}, \mu \in \mathbb{R}, \xi \in \mathbb{R^{+}}.
\label{tlsc}
\end{eqnarray}
We refer the distribution in Eq. (\ref{tlsc}) as the truncated-logistic-skew Cauchy (TLSC) distribution with parameters $\lambda, \mu$ and $\xi$.
In Figure 3, we plotted the pdfs of the TLSC distribution with $\mu = 0$ and $\xi = 1$ for different values of the parameter $\lambda$.

\item If $X \sim \text{Logistic}(\mu, s)$, i.e., $f_{X}(x)=\frac{\exp\left(- \frac{x - \mu}{s}\right)}{s \left[1 + \exp\left(- \frac{x - \mu}{s}\right) \right]^{2}}$ and
$F_{X}(x)= \frac{1}{1 + \exp\left(- \frac{x - \mu}{s}\right)}$, $x \in \mathbb{R}$, $\mu \in \mathbb{R}$ and $s \in \mathbb{R}^{+}$, in Definition 1, then Eq. (2.3) gives the pdf of the random variable $Y$ as
\begin{eqnarray}
\nonumber
f_Y(y; \mu, \xi, \lambda) & = & \bigg[\frac{2(1+e^{-\lambda})}{(1-e^{-\lambda})}\bigg]\\
\nonumber
& &\times \Bigg\{\frac{\lambda \exp\left(- \frac{y - \mu}{s} \right) \exp\left(- \frac{\lambda}{1 + e^{-(y-\mu)/s}} \right)}{ s
\left[1 +  \exp\left(- \frac{y - \mu}{s} \right)\right]^{2}
\left[1 +  \exp\left(- \frac{\lambda}{1 + e^{-(y-\mu)/s}} \right)\right]^{2}
} \Bigg\},\\
& & \qquad \qquad \qquad \qquad \qquad   \qquad y \in \mathbb{R}, \lambda\in \mathbb{R}, \mu \in \mathbb{R}, s \in \mathbb{R^{+}}.
\label{tlslg}
\end{eqnarray}
We refer the distribution in Eq. (\ref{tlslg}) as the truncated-logistic-skew Logistic (TLSLG) distribution with parameters $\lambda, \mu$ and $\xi$.
In Figure 4, we plotted the pdfs of the TLSLG distribution with $\mu = 0$ and $s = 1$ for different values of the parameter $\lambda$.

\end{enumerate}


\begin{figure}[htp]
 \begin{floatrow}
         \ffigbox{\includegraphics[scale = 0.42]{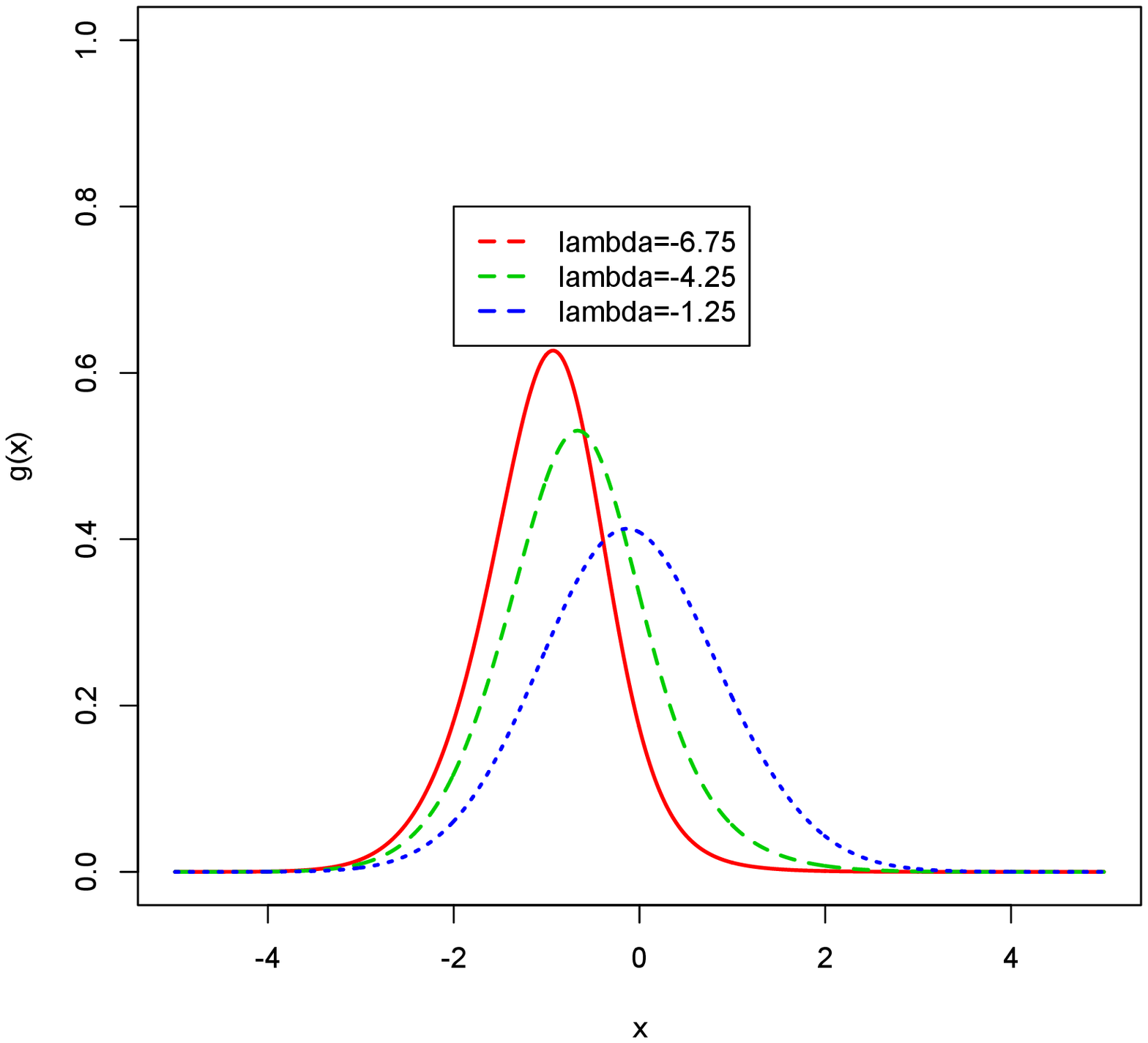}}{\caption{Probability density functions of different TLSN distributions}\label{case1}}
                   \ffigbox{\includegraphics[scale = 0.42]{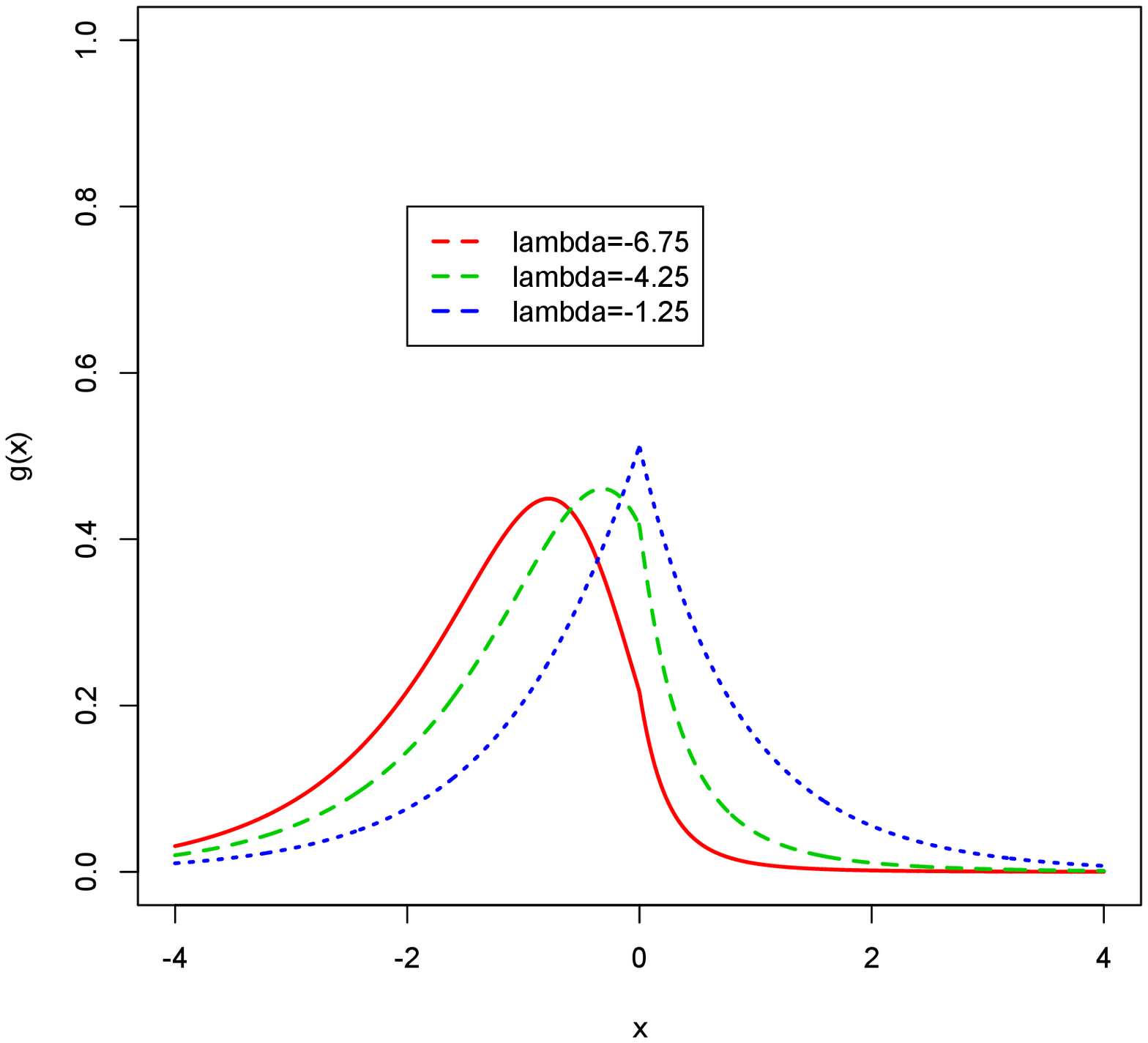}}{\caption{Probability density functions of different TLSL distributions}\label{zoom}}
                            \end{floatrow}
\end{figure}

\begin{figure}[htp]
 \begin{floatrow}
         \ffigbox{\includegraphics[scale = 0.42]{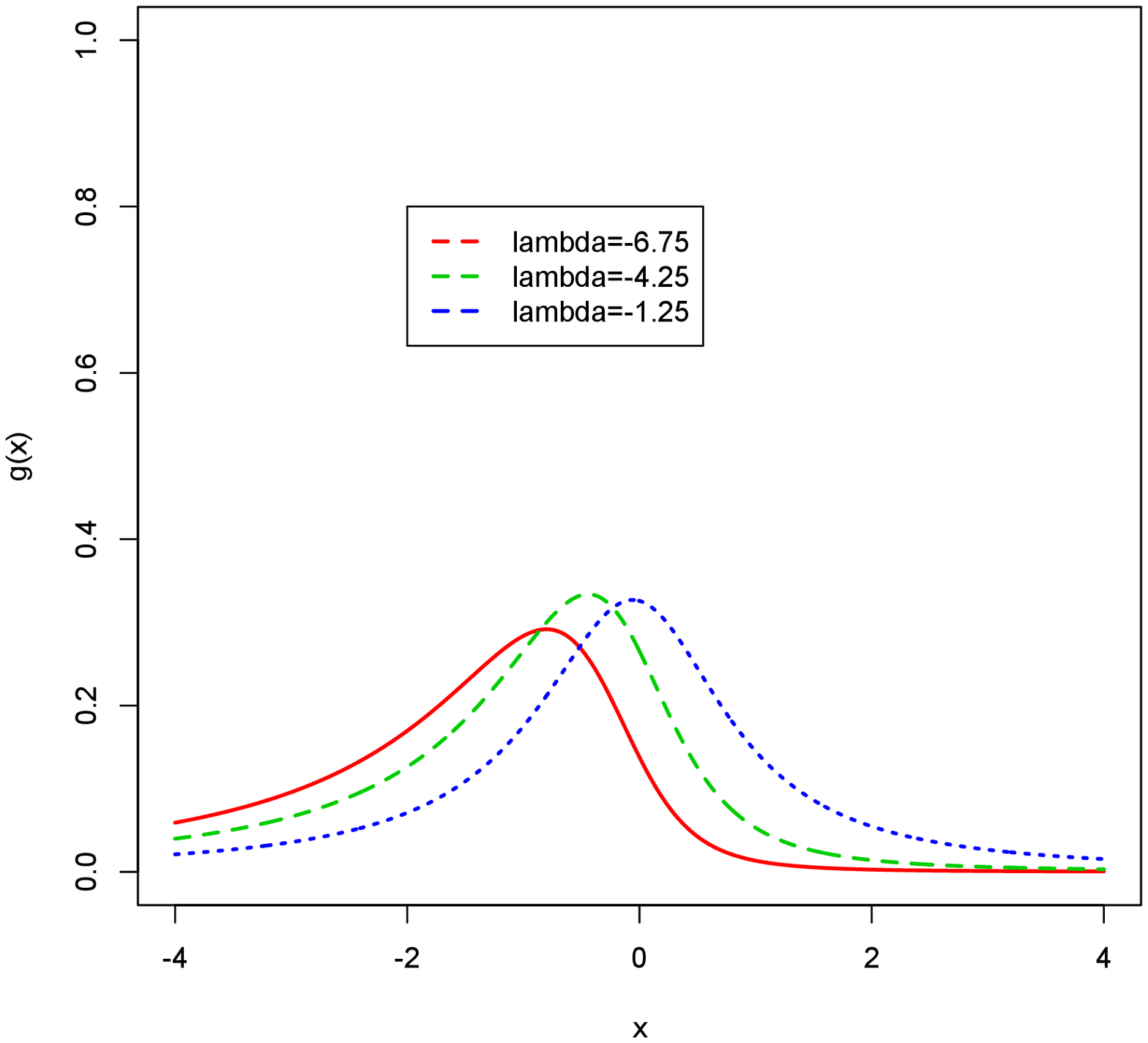}}{\caption{Probability density functions of different TLSC distributions}\label{case1}}
                   \ffigbox{\includegraphics[scale = 0.42]{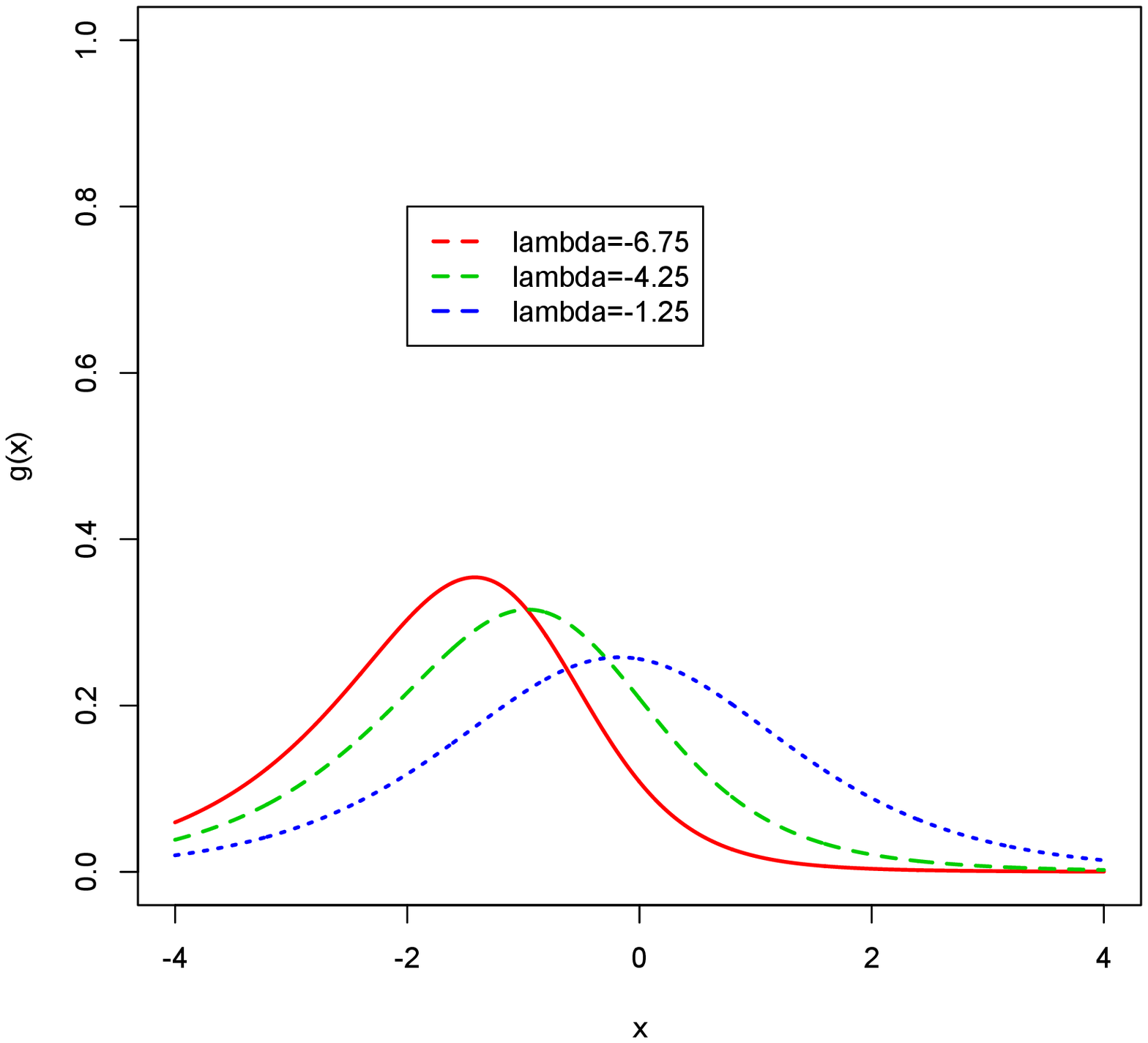}}{\caption{Probability density functions of different TLSLG distributions}\label{zoom}}
                            \end{floatrow}
\end{figure}

\newpage
\section{Structural Properties of the TLSS class of distributions}
\setcounter{equation}{0}
\label{S:2}

In this section, we study some important structural properties of the proposed class of skew-symmetric distribution.

\noindent
\textbf{Result 1:} Moment generating function and characteristic function: Let $M_{k:n}(t) = E[\exp(tX_{k:n})]$ and $\phi_{k:n}(t) = E[\exp(it X_{k:n})]$ denote the moment generating function (mgf) and the characteristic function (chf) of the $k$-th order statistic of a random sample of size $n$ from $F_X(\cdot)$, $X_{k:n}$, where $i=\sqrt{-1}$. Then, the mgf and chf of $Y \sim TLSS(\lambda)$ can be expressed as
\begin{eqnarray*}
E[\exp(tY)] & = & \frac{\lambda (1+e^{-\lambda})}{e^{-\lambda}} \sum_{j=0}^\infty \sum_{k=0}^\infty (-1)^{j+k+1} j \frac{[\lambda(j+1)]^k}{k!} M_{k+1:k+1}(t) \\
\mbox{ {and }} E[\exp(itY)] & = & \frac{\lambda (1+e^{-\lambda})}{e^{-\lambda}} \sum_{j=0}^\infty \sum_{k=0}^\infty (-1)^{j+k+1} j \frac{[\lambda(j+1)]^k}{k!} \phi_{k+1:k+1}(t),
\end{eqnarray*}
respectively.

\noindent
\textbf{Result 2:} Suppose $Y \sim TLSS(\lambda)$, if $ E(|X|^{r})$ exists for any $r \geq 1$, then $E(|Y|^{r})$ also exists.

\noindent
\textit{Proof.}  Note that
\begin{eqnarray}
 E (|Y|^{r}) & = & \int^{\infty}_{-\infty} \bigg(\frac{2(1+e^{-\lambda})}{(1-e^{-\lambda})}\bigg) \bigg[ \frac{\lambda f_x(y) e^{-\lambda F_x(y)}}{\{1+e^{-\lambda F_x(y)}\}^2} \bigg] |y|^r dy \nonumber \\
 & = & \bigg(\frac{2(1+e^{-\lambda})}{(1-e^{-\lambda})} \bigg) \cdot \lambda \int ^\infty _{-\infty} \mid Y \mid ^r \cdot \bigg \{ \frac{e^{-\lambda F_X(Y)}}{[1+e^{-\lambda F_X(Y)}]^2}\bigg\} f_X(y)dy  \nonumber \\
 & = & \frac{2\lambda (1+e^{-\lambda}}{(1-e^{-\lambda})} E \bigg(|X|^r \bigg\{ \frac{e^{-\lambda F_X(Y)}}{[1+e^{-\lambda F_X(Y)}]^2} \bigg\} \bigg).
\end{eqnarray}
For any real $Y \in (-\infty, \infty)$ and $\lambda > 0$, we have
\begin{eqnarray}
|X|^r  \frac{e^{-\lambda F_X(Y)}}{[1+e^{-\lambda F_X(Y)}]^2} \leq |X|^r,
\end{eqnarray}
since ${e^{-\lambda F_X(Y)}}/{[1+e^{-\lambda F_X(Y)}]^2}$ is always less than 1. Thus, $E (|Y|^r) \leq E(|X|^r) < \infty$. $\square$

\noindent
\textbf{Result 3:} Alternative expression for $E(Y^r)$:
Let $X_{k:n}$ denote the $k$-th order statistic from a random sample of size n from $F_X(\cdot)$ and $Y \sim TLSS(\lambda)$. If the conditions of Result 2 holds, then
\begin{equation}
E(Y^r) = \frac{2\lambda (1+e^{-\lambda})}{(1-e^{-\lambda})} \sum_{j=0}^\infty \sum_{k=0}^\infty (-1)^{j+k+1} j \cdot \frac{[\lambda(j+1)]^k}{k!} \cdot \frac{E[X^r_{k+1:k+1}]}{k+1}
\end{equation}

\noindent
\textit{Proof.} The $r$-th moment of the random variable $Y$ can be expressed as
\begin{eqnarray}
E(Y^r) & = & \bigg( \frac{2(1+e^{-\lambda})}{(1-e^{-\lambda})} \bigg) \cdot \lambda \int_{-\infty}^\infty y^r \bigg[ \frac{f_X(y)e^{-\lambda F_X(y)}}{ \{1+e^{-\lambda F_X(y)} \} ^2} \bigg] dy \nonumber \\
& = &\frac{2(1+e^{-\lambda})}{(1-e^{-\lambda})} \int_{-\infty}^\infty y^r \cdot f_X(y) e^{-\lambda F_X(y)} \bigg[\sum_{j=0}^\infty (-1)^{j+1} j \cdot e^{-\lambda j F_X(y)} \bigg] dy \nonumber \\
& = & \frac{2(1+e^{-\lambda})}{(1-e^{-\lambda})}\sum_{j=0}^\infty (-1)^{j+1} j \int_{-\infty}^\infty y^r \cdot f_X(y) e^{-\lambda (j+1) F_X(y)} dy \nonumber \\
& = & \frac{\lambda (1+e^{-\lambda})}{e^{-\lambda}} \sum_{j=0}^\infty (-1)^{j+1} j \int_{-\infty}^\infty y^r \cdot f_X(y) \bigg[ \sum_{k=0}^\infty (-1)^k \cdot \frac{\{ \lambda (J+1) \}^k}{k!} F_X^k(y) \bigg] dy \nonumber \\
& = &\frac{2(1+e^{-\lambda})}{(1-e^{-\lambda})} \sum_{j=0}^\infty \sum_{k=0}^\infty (-1)^{j+k+1} \cdot j \cdot \frac{[\lambda(j+1)]^k}{k!} \int_{-\infty}^\infty y^r f_X(y)F_X^k(y)dy \nonumber \\
& = & \frac{2(1+e^{-\lambda})}{(1-e^{-\lambda})} \sum_{j=0}^\infty \sum_{k=0}^\infty (-1)^{j+k+1}j \cdot  \frac{[\lambda(j+1)]^k}{k! (k+1)} E(X^r_{k+1:k+1}). \qquad \square
\end{eqnarray}
This alternative expression of $E(Y^r)$ can be used to obtain the mgf and chf discussed in Result 1.

\noindent
\textbf{Result 5:} Tail behavior property of TLSS($\lambda$): First, note that $Y \sim TLSS (\lambda)$ and $X$ is a symmetric random variable about $0$ with the cdf and pdf as $F_{X}(x)$ and $f_{X}(x)$ respectively. Then, the tails of $Y$ have the same behavior as the tails of $X$ because
\begin{eqnarray*}
f_Y(y) & \sim & \frac{2\lambda(1+e^{-\lambda})}{(1-e^{-\lambda})} \cdot \frac{e^{-\lambda}}{(1+e^{-\lambda})^{2}} f_X(y) = \frac{2\lambda e^{-\lambda} f_X(y)}{1+e^{-2\lambda}}, \quad \mbox{as} \quad y \rightarrow +\infty, \\
f_Y(y) & \sim & \bigg(\frac{2(1+e^{-\lambda})}{(1-e^{-\lambda})} \bigg) f_X(y), \quad \mbox{as} \quad y \rightarrow -\infty, \\
F_Y(y) & \sim & \frac{2\lambda(1+e^{-\lambda})}{(1-e^{-\lambda})} F_X(y), \quad \mbox{as} \quad y \rightarrow -\infty, \\
\mbox{{and }} 1-F_Y(y) & \sim & \frac{2\lambda e^{-\lambda}}{1-e^{-2\lambda}} \big(1-F_X(y)\big) \quad \mbox{as} \quad y \rightarrow +\infty.
\end{eqnarray*}

\noindent
\textbf{Result 6:} Mode: The mode of the random variable $Y \sim$ TLSS($\lambda$) can be obtained by taking the first-order derivative of the density function and subsequently equating it to zero:
\begin{eqnarray*}
&  & \frac{d}{dy} f_Y(y) = 0 \nonumber \\
& \Rightarrow & \lambda \bigg[\frac{2(1+e^{-\lambda})}{(1-e^{-\lambda})} \bigg] \bigg\{ \frac{[1+e^{-\lambda F_X(y)}]^2 [f_X'(y) e^{-\lambda F_X(y)} - \lambda f^2_X(y) e^{-\lambda F_X(y)}] - A_1}{[1+e^{-\lambda F_X(y)}]^4} \bigg\} = 0,
\end{eqnarray*}
where $A_1 = -2 \lambda [f_X(y) e^{-\lambda F_X(y)}]^{2} [1+e^{-\lambda F_X(y)}]$. After some algebraic simplification we obtain the following equation:
\begin{eqnarray}
& & [1+e^{-\lambda F_X(y)}] \bigg\{ [f'_X(y) - \lambda f^2_X(y)]  + 2\lambda f^2_X(y) e^{-\lambda F_X(y)} \bigg\} = 0 \nonumber \\
& \Rightarrow & [1+e^{-\lambda F_X(y)}] \cdot f'_X(y) + \lambda f^2_X(y) [1-e^{-\lambda F_X(y)}] = 0 \nonumber \\
& \Rightarrow & \frac{f'_X(y)}{f^2_X(y)} = \frac{\lambda[e^{-\lambda F_X(y)} - 1]}{[1+e^{-\lambda F_X(y)}]}.
\label{mode}
\end{eqnarray}
The roots of Eq. (\ref{mode}) are the modes of the random variable $Y \sim$  TLSS($\lambda$). Note that the roots are to the left (right) of zero for $\lambda > 0$ ($\lambda < 0$).
The root of Eq. (\ref{mode}), say $y=y_0$, corresponds to a maximum if,
\begin{eqnarray}
\frac{d}{dy} \bigg[\frac{d}{dy} f_Y(y) \bigg] \bigg|_{y=y_0} &<& 0 \nonumber \\
\Leftrightarrow f''_X(y_0) [1+e^{-\lambda F_X(y_0)}] + \lambda^2 f^3_X(y_0)e^{-\lambda F_X(y_0)} &<& \lambda f_X(y_0) f'_X(y_0) \{3e^{-\lambda F_X(y_0)} - 2 \} \nonumber
\end{eqnarray}
Similarly, the root of Eq. (\ref{mode}), $y_0$, corresponds to a minimum if,
\begin{eqnarray}
f''_X(y_0) [1+e^{-\lambda F_X(y_0)}] + \lambda^2 f^3_X(y_0)e^{-\lambda F_X(y_0)} & > & \lambda f_X(y_0) f'_X(y_0) \{3e^{-\lambda F_X(y_0)} - 2 \}. \nonumber
\end{eqnarray}
The root of Eq. (\ref{mode}) corresponds to a inflection point if,
\begin{eqnarray}
f''_X(y_0) [1+e^{-\lambda F_X(y_0)}]+ \lambda ^2f^3_x(y_0)e^{-\lambda F_X(y_0)} & = & \lambda f_X(y_0) f'_x(y_0) \{3e^{-\lambda F_X(y_0)} - 2 \}. \nonumber
\end{eqnarray}
The mode corresponding to a maximum is unique if $y_0$ satisfies
\begin{eqnarray*}
& & f'_X(y) > f^2_X(y) \cdot \bigg\{\frac{(e^{-\lambda F_X(y)} - 1) \lambda}{[1+e^{-\lambda F_X(y)}]}\bigg\} \qquad \mbox{for all} \quad y<y_0,\\
& {\mbox {and }} & f'_X(y) < f^2_X(y) \cdot \bigg\{\frac{(e^{-\lambda F_X(y)} - 1) \lambda}{[1+e^{-\lambda F_X(y)}]}\bigg\} \qquad \mbox{for all} \quad y > y_0.
\end{eqnarray*}
Similarly, the mode corresponding to a minimum is unique if $y_0$ satisfies
\begin{eqnarray*}
&                & f'_X(y) < f^2_X(y) \cdot \bigg\{\frac{(e^{-\lambda F_X(y)} - 1) \lambda}{[1+e^{-\lambda F_X(y)}]} \bigg\} \qquad \mbox{for all} \quad y<y_0,\\
& {\mbox {and }} & f'_X(y) > f^2_X(y) \cdot \bigg\{\frac{(e^{-\lambda F_X(y)} - 1) \lambda}{[1+e^{-\lambda F_X(y)}]}\bigg\} \qquad \mbox{for all} \quad y > y_0.
\end{eqnarray*}

\section{Generating Random Variates from the TLSS class of distributions}
\setcounter{equation}{0}

In this section, we discuss the generation of the random variates from the TLSS distribution based on the inverse transform method and an acceptance-rejection method. Since the cdf in Eq. (\ref{cdf}) of the random variable follows the TLSS distribution is continuous, the cdf $F_{Y}$ is invertible  with the inverse cdf presented in Eq. (\ref{icdf}). Based on the inverse transform method, a random variate from the TLSS distribution with specific value of $\lambda$ and $F_{X}$ can be generated by the following steps:
\begin{itemize}
\item[Step 1.] Generate a random variate $U$ from the uniform distribution in (0, 1), i.e., $U \sim U(0, 1)$.
\item[Step 2.] Obtain the random variate $Y$ by solving $Y = F_{Y}^{-1}(U; \lambda)$, where $F_{Y}^{-1}(\cdot; \lambda)$ is presented in Eq. (2.3).
\end{itemize}

\noindent In general, there is no closed form solution for the equation $y = F_{Y}^{-1}(u; \lambda)$ and hence, numerical method is required to solve the non-linear equation in order to obtain the random variate $y$ that follows the TLSS distribution. To avoid using a numerical method, we consider the acceptance-rejection method by using $F_{X}$ as the proposed distribution. 
The acceptance-rejection method provides an alternative way to generate $Y \sim TLSS(\lambda)$ if $f_{X}$ is a density that can easily be simulated from. The following acceptance-rejection algorithm can be used to generate $Y \sim TLSS(\lambda)$:
\begin{itemize}
\item[Step 1.] Generate $X \sim f_{X}$ as a proposal.
\item[Step 2.] Generate $U \sim U(0, 1)$.
\item[Step 3.] If $U \leq \frac{4 \exp[-\lambda f_{x}(X)]}{\left\{1 + \exp[-\lambda f_{x}(X)] \right\}^{2}}$, set $Y = X$. Otherwise, return to Step 1.
\end{itemize}

\section{Estimation of Model Parameters}
\setcounter{equation}{0}

\subsection{Maximum Likelihood Estimators and Fisher Information}

Suppose that $\y = (y_{1}, y_{2}, \ldots, y_{n})$ is a random sample of size $n$ from the distribution with pdf in Eq. (\ref{pdf}) and $\btheta$ is the parameter vector of the symmetric distribution $f_{X}$, then the log-likelihood equation can be written as
\begin{eqnarray}
\nonumber
\ln L(\lambda,\btheta) & = & \sum\limits_{i=1}^{n} \ln f(y_i; \lambda, \btheta) \\
\nonumber
& = & n \ln 2 + n \ln \lambda + n \ln (1 + e^{-\lambda}) - n \ln (1 - e^{-\lambda}) \\
&   & + \sum\limits_{i=1}^{n} \ln f_{X}(y_{i}; \btheta) - \lambda \sum\limits_{i=1}^{n} F_{X}(y_{i}; \btheta) - 2 \sum\limits_{i=1}^{n} \ln\left[1 + e^{-\lambda F_{x}(y_{i}; \btheta)} \right].
\label{llik}
\end{eqnarray}
The maximum likelihood estimator (MLE) of $\btheta$, denoted as $\hbtheta$, can be obtained by maximizing the log-likelihood function in Eq. (\ref{llik}) with respect to $\btheta$.
Under standard regularity conditions, as $n\rightarrow \infty$, the distribution of $\hbtheta$ can be approximated by a multivariate normal distribution $N_{p+1}(\btheta, {\bf J}(\hbtheta)^{-1})$, where $p$ is the number of parameters in the distribution $f_{X}$. Here, ${\bf J}(\hbtheta)$ is the observed information matrix evaluated at the maximum likelihood estimate $\hbtheta$.

\noindent For illustrative purpose, we consider the TLSN distribution in Eq. (\ref{tlsn}) with the log-likelihood function
\begin{eqnarray}
\nonumber
\ell(\lambda,\mu,\sigma) & = & \ln L (\lambda,\mu,\sigma) \\
\nonumber
& = & n\left(\ln 2+\ln \lambda\right) + n  \ln (1+e^{-\lambda}) - n  \ln (1-e^{-\lambda}) + \sum^n_{i=1} \ln \bigg \{ \frac{\phi(\frac{y_i-\mu}{\sigma}) e^{-\lambda \Phi(\frac{y_i-\mu}{\sigma})}}{ \{1+e^{-\lambda \Phi(\frac{y_i-\mu}{\sigma})} \}^2} \bigg\}  \\
& = & -n\ln \sigma+n\left(\ln 2+\ln \lambda\right)+n  \ln (1+e^{-\lambda}) - n  \ln (1-e^{-\lambda}) \nonumber\\
& & + \sum^n_{i=1} \ln \phi\left(\frac{y_i-\mu}{\sigma}\right) - \lambda \sum^n_{i=1} \Phi\left(\frac{y_i-\mu}{\sigma}\right) - 2 \sum^n_{i=1} \ln \{1+e^{-\lambda \Phi\left(\frac{y_i-\mu}{\sigma}\right)} \}.
\label{tlsnlik}
\end{eqnarray}
The MLEs of the parameters in the TLSN distribution, $\lambda$, $\mu$, $\sigma$, can be obtained by taking the partial derivatives of $\ell(\lambda,\mu,\sigma)$ with respect to $\lambda$, $\mu$ and $\sigma$ respectively and set them to zero. We have the maximum likelihood equations:
\begin{eqnarray}
\label{leq1}
\frac{\partial \ell}{\partial \lambda}
& = & \frac{n}{\lambda}-\frac{2n\lambda}{(1+e^{-2\lambda})}-\sum_{i=1}^{n} \Phi\left(\frac{y_i-\mu}{\sigma}\right)
+2\sum_{i=1}^{n}\frac{\Phi(\frac{y_i-\mu}{\sigma}) e^{-\lambda \Phi(\frac{y_i-\mu}{\sigma})}}{ \{1+e^{-\lambda \Phi(\frac{y_i-\mu}{\sigma})} \}} = 0,\\
\frac{\partial \ell}{\partial \mu}
\label{leq2}
& = & -\sum_{i=1}^{n}\frac{1}{\sigma\phi(\frac{y_i-\mu}{\sigma})}+\sum_{i=1}^{n}\frac{\lambda \phi(\frac{y_i-\mu}{\sigma})}{\sigma}
-2\sum_{i=1}^{n}\frac{\phi(\frac{y_i-\mu}{\sigma})e^{-\lambda \Phi(\frac{y_i-\mu}{\sigma})}}{\sigma  \{1+e^{-\lambda \Phi(\frac{y_i-\mu}{\sigma})} \}} = 0,\\
\nonumber
\frac{\partial \ell}{\partial \sigma} & = & -\frac{n}{\sigma} + \frac{\mu}{\sigma^{2}} \left\{ \sum_{i=1}^{n}\left[\phi(\frac{y_i-\mu}{\sigma})\right]^{-1}
- \lambda\sum_{i=1}^{n}\phi\left(\frac{y_i-\mu}{\sigma}\right) \right.\\
& & \left. -2\lambda\sum_{i=1}^{n}\phi\left(\frac{y_i-\mu}{\sigma}\right)e^{-\lambda \Phi(\frac{y_i-\mu}{\sigma})}[1+e^{-\lambda \Phi(\frac{y_i-\mu}{\sigma})}]^{-1}\right\} = 0.
\label{leq3}
\end{eqnarray}
Solving Eqs. (\ref{leq1})--(\ref{leq3}) for $\lambda, \mu$ and $\sigma$ simultaneously gives the maximum likelihood estimates of $\lambda, \mu$ and $\sigma$, denoted as ${\hat \lambda}$, ${\hat \mu}$ and ${\hat \sigma}$, respectively. Here, the observed Fisher information matrix ${\bf J}(\hbtheta)$ is given by
\[
{\bf J}(\hbtheta)=
  \begin{bmatrix}
    J_{\lambda \lambda } & J_{\lambda \mu } &J_{\lambda \sigma }\\
       & J_{\mu \mu } & J_{\mu \sigma }\\
       &   &  J_{\sigma \sigma }
  \end{bmatrix},
\]
where 
\begin{eqnarray*}
J_{\lambda \lambda} & = & \left. - \frac{\partial^2 \ell}{\partial \lambda^{2}} \right\vert_{\btheta = \hbtheta}\\
& = & \frac{n}{\lambda^{2}} -\frac{2n}{\left(1+e^{-2\lambda}\right)^{2}} \left[1+e^{-2\lambda} (1+2\lambda)\right]
 +2\sum_{i=1}^{n}e^{-2\lambda\Phi\left(\frac{y_i-\mu}{\sigma} \right)} \left[\frac{ \Phi\left(\frac{y_i-\mu}{\sigma}\right)}{1+e^{-\lambda\Phi\left(\frac{y_i-\mu}{\sigma}\right)}}\right]^{2},  \\
\end{eqnarray*}

\begin{eqnarray*}
J_{\mu\mu} & = & \left. -\frac{\partial^{2} \ell}{\partial \mu^{2}}\right\vert_{\btheta = \hbtheta}\\
& = & \frac{1}{\sigma^{2}} \left\{ \sum_{i=1}^{n} \frac{\phi^{(1)}\left(\frac{y_i-\mu}{\sigma}\right)}{\left[\phi\left(\frac{y_i-\mu}{\sigma}\right)\right]^{2}} + \lambda\sum_{i=1}^{n}\phi^{(1)}\left(\frac{y_i-\mu}{\sigma}\right) \right\}\\
& & - \frac{2}{\sigma^{2}} \sum_{i=1}^{n} \left\{ \left[1+e^{-\lambda\Phi \left(\frac{y_i-\mu}{\sigma} \right)}\right]^{-2} e^{-\lambda\Phi\left(\frac{y_i-\mu}{\sigma}\right)} \right.\\
& & \qquad \qquad \times \left. \left[ \phi^{(1)}\left(\frac{y_i-\mu}{\sigma}\right) + \lambda \phi^{2}\left(\frac{y_i-\mu}{\sigma}\right)
+\phi^{(1)}\left(\frac{y_i-\mu}{\sigma} \right)e^{-\lambda \Phi\left(\frac{y_i-\mu}{\sigma} \right)} \right] \right\},\\
J_{\lambda \mu} & = & \left. -\frac{\partial^{2} \ell}{\partial\lambda\partial \mu}\right\vert_{\btheta = \hbtheta}\\
& = & - \frac{1}{\sigma} \sum_{i=1}^{n} \phi\left(\frac{y_i-\mu}{\sigma} \right)\\
&   & + \frac{2}{\sigma} \sum_{i=1}^{n} \phi\left(\frac{y_i-\mu}{\sigma}\right) 
\left[1 + e^{-\lambda\Phi\left(\frac{y_i-\mu}{\sigma}\right)}\right]^{-2} \Phi\left(\frac{y_i-\mu}{\sigma}\right)
\left[1 + 2 e^{-2\lambda\Phi(\frac{y_i-\mu}{\sigma})}\right],\\
J_{\lambda \sigma} & = & \left. -\frac{\partial^{2} \ell}{\partial\lambda\partial \sigma}\right\vert_{\btheta = \hbtheta}\\
& = & -\frac{\mu}{\sigma^{2}} \sum_{i=1}^{n}\phi(\frac{y_i-\mu}{\sigma}) + \frac{2\mu}
{\sigma^{2}} \left[\sum_{i=1}^{n}\phi\left(\frac{y_i-\mu}{\sigma} \right) e^{-\lambda\Phi\left(\frac{y_i-\mu}{\sigma}\right)}
\left[1+e^{-\lambda\Phi(\frac{y_i-\mu}{\sigma})}\right]^{-1}\right.\\
& &  -2\lambda\sum_{i=1}^{n}\phi\left(\frac{y_i-\mu}{\sigma}\right) 
\left[1 + e^{-\lambda\Phi\left(\frac{y_i-\mu}{\sigma}\right)}\right]^{-2} \Phi\left(\frac{y_i-\mu}{\sigma}\right)
\left[1 + 2 e^{-2\lambda\Phi(\frac{y_i-\mu}{\sigma})}\right] e^{-\lambda\Phi(\frac{y_i-\mu}{\sigma})},\\
J_{\sigma \mu} & = & \left. -\frac{\partial^{2} \ell}{\partial \sigma\partial \mu}\right\vert_{\btheta = \hbtheta} \\
& = & -\sum_{i=1}^{n}\frac{\phi(\frac{y_i-\mu}{\sigma})\phi^{(2)}\left(\frac{y_i-\mu}{\sigma}\right)}{\left[\sigma\phi\left(\frac{y_i-\mu}{\sigma}\right)\right]^{2}} 
-\lambda \sum_{i=1}^{n}\frac{\sigma \phi^{(2)}\left(\frac{y_i-\mu}{\sigma}\right)-\phi\left(\frac{y_i-\mu}{\sigma}\right)}{\sigma^{2}} +2 \sum_{i=1}^{n}\frac{M_{1}-M_{2}}{\sigma^{2} \left[1+e^{-\lambda\Phi\left(\frac{y_i-\mu}{\sigma}\right)} \right]^{2}},\\
J_{\sigma \sigma} & = & \left. -\frac{\partial^{2} \ell}{\partial \sigma^{2}}\right\vert_{\btheta = \hbtheta} = -\frac{n}{\sigma^{2}}+\frac{2\mu B}{\sigma^{3}}-\frac{\mu}{\sigma^{2}} \left( \frac{\partial B}{\partial\sigma} \right),
\end{eqnarray*}
with 
\begin{eqnarray*}
\phi^{(1)}(\frac{y_i-\mu}{\sigma}) & = & \frac{\partial}{\partial\mu} \phi(\frac{y_i-\mu}{\sigma}), \\
\phi^{(2)}(\frac{y_i-\mu}{\sigma}) & = & \frac{\partial}{\partial\sigma} \phi(\frac{y_i-\mu}{\sigma}),\\ 
M_{1} & = & \{1+e^{-\lambda\Phi(\frac{y_i-\mu}{\sigma})}\}e^{-\lambda\Phi(\frac{y_i-\mu}{\sigma})} \left[\phi^{(2)}(\frac{y_i-\mu}{\sigma})-\frac{\phi^{2}(\frac{y_i-\mu}{\sigma})}{\sigma}\right],\\
M_{2} & = & \phi(\frac{y_i-\mu}{\sigma})e^{-\lambda\Phi(\frac{y_i-\mu}{\sigma})} \left\{1+e^{-\lambda\Phi(\frac{y_i-\mu}{\sigma})}\left[1-\lambda\sigma \phi(\frac{y_i-\mu}{\sigma})\right]\right\}, \\
B & = & \sum_{i=1}^{n} \left[\phi\left(\frac{y_i-\mu}{\sigma}\right) \right]^{-1} -\lambda\sum_{i=1}^{n}\phi\left(\frac{y_i-\mu}{\sigma} \right)\\
& & -2\lambda\sum_{i=1}^{n}\phi\left(\frac{y_i-\mu}{\sigma} \right) e^{-2\lambda\Phi\left(\frac{y_i-\mu}{\sigma}\right)} \left[1+e^{-\lambda\Phi\left(\frac{y_i-\mu}{\sigma} \right)}\right]^{-1},\\
\frac{\partial B}{\partial\sigma} & = & \frac{\mu}{\sigma^{2}} \sum_{i=1}^{n} \left[ \phi\left(\frac{y_i-\mu}{\sigma} \right)\right]^{-2} \phi^{(2)}\left(\frac{y_i-\mu}{\sigma} \right)
+\frac{\mu\lambda}{\sigma^{2}}\sum_{i=1}^{n}\phi^{(2)}\left(\frac{y_i-\mu}{\sigma}\right) \\
& & +\frac{2\lambda\mu}{\sigma^{2}}\sum_{i=1}^{n}\left[\phi^{(2)}\left(\frac{y_i-\mu}{\sigma}\right) \left[ 1 + e^{-\lambda \Phi\left(\frac{y_i-\mu}{\sigma}\right)}\right]^{-1}e^{-\lambda\Phi(\frac{y_i-\mu}{\sigma})}\right. \\
& & \left.  \qquad \qquad \qquad -\phi^{2}\left(\frac{y_i-\mu}{\sigma} \right)\left[1+e^{-\lambda\Phi\left(\frac{y_i-\mu}{\sigma} \right)}\right]^{-1} e^{-\lambda\Phi(\frac{y_i-\mu}{\sigma})}\right].
\end{eqnarray*}


\noindent The asymptotic variance-covariance matrix of the MLE $\hbtheta = ({\hat \theta}_{1}, {\hat \theta}_{2}, {\hat \theta}_{3}) = ({\hat \mu}, {\hat \sigma}, {\hat \lambda})$ can be obtained from the inverse of the observed Fisher information matrix as
$$ {\bf V} = {\bf J}^{-1}(\hbtheta) =
\begin{bmatrix}
v_{11} & v_{12} & v_{13}\\
       & v_{22} & v_{23}\\
       &        & v_{33}
  \end{bmatrix}.
$$
Then, based on the asymptotic normality of the MLE, a $100(1-\delta)\%$ approximate confidence interval for the parameter $\theta_{i}$ can be obtained as
\begin{eqnarray}
{\hat \theta}_{i} \pm z_{1 - \delta/2} \sqrt{v_{ii}},
\label{naci}
\end{eqnarray}
where $z_{q}$ is the 100$q$-th upper percentile of the standard normal distribution.

\subsection{Simulation study}

In this subsection, we perform a Monte Carlo simulation study to evaluate the performance of the likelihood inference for the TLSN distribution in Eq. (\ref{tlsn}). We consider the sample sizes $n$ = 50, 75 and 100 with parameters $\mu = 0$, $\sigma = 1$ and different values of $\lambda = -1.5$, -1, 1 and 1.5. Random samples from the TLSN distribution are generated from the acceptance-rejection method presented in Section 4. The MLEs of $\mu$, $\sigma$ and $\lambda$ are obtained by maximizing the likelihood function in Eq. (\ref{tlsnlik}) with the {\tt optim} function in R (R Core Team, 2018). For each simulated random sample, we also compute the 95\% approximate confidence intervals based on Eq. (\ref{naci}). For each setting, $10000$ set of random samples are generated.

The estimated biases and mean squared errors of the MLEs of $\mu$, $\sigma$ and $\lambda$ are presented in Table \ref{biasmse}. The estimated coverage probabilities and average widths are presented in Table \ref{cpaw}. Since the observed information need not be positive definite which results in negative asymptotic variances (see, for example, Verbeke and Molenberghs, 2007), we also presented the percentage of cases in which the asymptotic variances are negative and the confidence intervals cannot be computed in Table \ref{cpaw}.

\begin{table}[htp]
\begin{center}
{
\caption{Simulated biases and mean squared errors (MSEs)  of the MLEs of the parameters in the TLSN distribution with $\mu = 0$, $\sigma = 1$ and different values of $\lambda$}
\label{cpaw}
\begin{tabular}{ c  c c c c c c c } \hline
          &     & \multicolumn{2}{c}{$\mu$} &  \multicolumn{2}{c}{$\sigma$}  & \multicolumn{2}{c}{$\lambda$}  \\
$\lambda$ & $n$ & Bias & MSE & Bias & MSE & Bias & MSE \\ \hline
1   & 50  & 0.131 & 0.143 & 0.021 & 0.019  & 0.316  & 3.263\\
    & 75  & 0.137 & 0.144 & 0.026 & 0.017  & 0.340  & 3.192\\
    & 100 & 0.126 & 0.122 & 0.029 & 0.013  & 0.302  & 2.798\\ \hline

1.5 & 50  & 0.089 & 0.156 & 0.014 & 0.021  & 0.033  & 4.590\\
    & 75  & 0.093 & 0.142 & 0.021 & 0.018  & 0.048  & 3.498\\
    & 100 & 0.090 & 0.129 & 0.022 & 0.015  & 0.043  & 3.098\\ \hline

-1  & 50  & 0.130 & 0.145 & 0.020 & 0.019  & -0.294 & 3.216\\
    & 75  & 0.132 & 0.140 & 0.027 & 0.017  & -0.311 & 3.110\\
    & 100 & 0.126 & 0.126 & 0.027 & 0.014  & -0.286 & 2.842\\ \hline

-1.5 & 50 & 0.085 & 0.151 & 0.013 & 0.020  & -0.013 & 4.433\\
     & 75 & 0.091 & 0.138 & 0.020 & 0.017  & -0.033 & 3.338\\
    & 100 & 0.089 & 0.128 & 0.022 & 0.015  & -0.035 & 3.065\\ \hline
\end{tabular}
}
\end{center}
\end{table}

\begin{table}[htp]
\begin{center}
{
\caption{Simulated coverage probabilities (CP) and average widths (AW) of the MLEs of the parameters in the TLSN distribution with $\mu = 0$, $\sigma = 1$ and different values of $\lambda$}
\label{biasmse}
\begin{tabular}{ c  c c c c c c c c} \hline
          &     & \multicolumn{2}{c}{$\mu$} &  \multicolumn{2}{c}{$\sigma$}  & \multicolumn{2}{c}{$\lambda$} & \\
$\lambda$ & $n$ & CP & AW & CP & AW  & CP & AW  & \% of CI cannot be computed \\ \hline
1 & 50 & 0.950 & 1.377 & 0.952 & 0.563 & 0.992 & 12.361 & 0.060 \\
  & 75 & 0.939 & 1.243 & 0.953 & 0.487 & 0.986 & 11.261 & 0.090 \\
 & 100 & 0.935 & 1.137 & 0.959 & 0.432 & 0.982 & 10.479 & 0.110 \\ \hline
1.5 & 50 & 0.905 & 1.444 & 0.940 & 0.577 & 0.997 & 12.596 & 0.130 \\
    & 75 & 0.882 & 1.292 & 0.943 & 0.498 & 0.993 & 11.189 & 0.170 \\
   & 100 & 0.853 & 1.203 & 0.943 & 0.448 & 0.988 & 10.190 & 0.090 \\ \hline
-1 & 50 & 0.950 & 1.371 & 0.949 & 0.561 & 0.993 & 12.480 & 0.110 \\
   & 75 & 0.941 & 1.229 & 0.956 & 0.484 & 0.986 & 11.264 & 0.170 \\
   & 100 & 0.930 & 1.117 & 0.955 & 0.428 & 0.978 & 10.262 & 0.110 \\ \hline
-1.5 & 50 & 0.904 & 1.437 & 0.940 & 0.575 & 0.997 & 12.556 & 0.130 \\
     & 75 & 0.882 & 1.289 & 0.943 & 0.496 & 0.993 & 11.219 & 0.090 \\
    & 100 & 0.853 & 1.199 & 0.943 & 0.447 & 0.988 & 10.199 & 0.100  \\ \hline
\end{tabular}
}
\end{center}
\end{table}


From the simulation results in Table 1, the estimated MSEs for the three parameters $\mu$, $\sigma$ and  $\lambda$ decreases as the sample size increases, which is a desirable property for an efficient estimator. However, for the estimated biases, there is not a steady decreasing pattern with the increase of sample sizes. 
We can observe that the direction of the estimated biases for the MLE of $\lambda$ is the same as the sign of the true value of the parameter $\lambda$. Moreover, the estimated MSEs of $\lambda$ is larger than the MSEs of $\mu$ and $\sigma$. This is not totally unexpected since the estimation of the shape parameter for skewed probability models are known to be challenging even for large sample sizes (see, for example, Pewsey, 2000). 
        
From Table 2, we can see that the proportions of cases that the estimated asymptotic variances being negative are very small ($< 0.2\%$), which can be  negligible. We can also observe that when the sample size increases, the estimated average widths of the confidence intervals get smaller. For the coverage probabilities of the confidence intervals, the estimated coverage probabilities of the confidence intervals for $\sigma$ and $\lambda$ are always close to or above the nominal level 95\% in all the settings considered here, however, the estimated coverage probabilities of the confidence intervals for $\mu$ can be lower than the nominal level, which indicates that one should be caution when using the approximate confidence interval for $\mu$.  

\section{Applications in Environmental Studies}
\setcounter{equation}{0}

\subsection{Mercury concentrations in swordfish}

Lee and Krutchkoff (1980) presented the actual mercury concentrations found in 115 swordfish. They assumed that the mercury concentration $X$ has a two-parameter lognormal distribution, which is a skewed distribution, and studied the relationship between the mean and variance. The two-parameter lognormal distribution considered in Lee and Krutchkoff (1980) has a pdf
\begin{eqnarray}
f(y; \mu^{*}, \sigma^{*}) & = & \frac{1}{\sigma^{*}\sqrt{2 \pi}} \exp \left[- \frac{(y - \mu^{*})^{2}}{2\sigma^{*2}} \right], y \in \mathbb{R^{+}}, \mu^{*} \in \mathbb{R}, \sigma^{*} \in \mathbb{R^{+}}.
\label{lnpdf}
\end{eqnarray}
Based on the $115$ observations presented in Table $1$ of Lee and Krutchkoff (1980), the MLEs of the model parameters $\mu^{*}$ and $\sigma^{*}$ in the lognormal distribution are  ${\hat \mu}^{*} = -0.0688$ and  $\sigma^{*} = 0.7025$, respectively, which gives the maximum log-likelihood as -114.17.

\begin{figure}[htp]
\includegraphics[scale = .6]{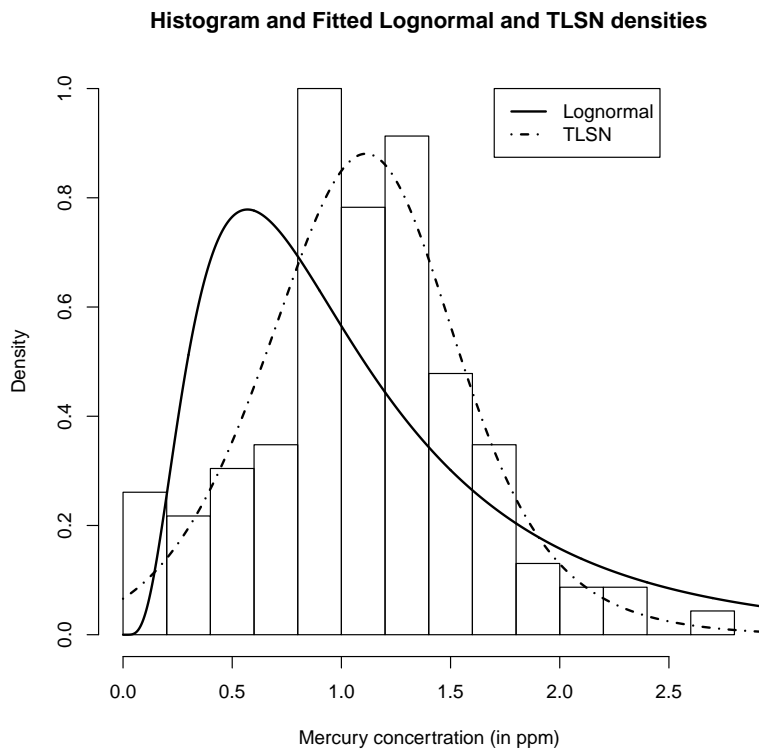}
{\caption{Histogram of the mercury concentrations in swordfish data and the fitted densities of lognormal and TLSN distributions}
\label{eg1}}
\end{figure}

Here, we propose the use of the TLSN distribution in Eq. (\ref{tlsn}) to fit the mercury concentrations data. Based on the expressions presented in Section 5.1, we can obtain the MLEs of the parameters in the TLSN distribution as
${\hat \mu} = 1.5585$ ${\hat \sigma} = 0.6221$ and ${\hat \lambda} = 4.6496$, which gives the maximum log-likelihood as -82.07. Since the lognormal model has one less estimated parameter compared to the TLSN distribution, to compare the relative fitting of the two statistical models for the mercury concentrations data, we consider the Akaike information criterion (AIC) defined as
$$AIC = 2k - 2 \ln {\hat L},$$
where $k$ is the number of estimated parameters in the model and $\hat L$ be the maximum value of the likelihood function for the model. The AIC values for the lognormal model and the TLSN model are, respectively, 232.34 and 170.14, which indicates that the TLSN model provides a better goodness-of-fit for the mercury concentrations data. To further illustrate the advantage of the TLSN model for fitting the mercury concentrations data, we plotted the histogram of the mercury concentrations data with the fitted pdfs of the lognormal and TLSN distributions in Figure \ref{eg1}. We can see that the TLSN provides a much better fit to the data compared to the lognormal distribution in the price of an extra model parameter.

For illustrative purposes, we also compute the 95\% approximated confidence intervals for $\mu$, $\sigma$ and $\lambda$ based on the variance-covariance matrix obtained from the inverse of the observed Fisher information matrix presented in Section 5.1, respectively, as (0.91627, 2.20143), (0.38518, 0.85916) and (-0.30235, 9.60490).

\subsection{Ammonium concentration in precipitation}

In this subsection, we consider an data set from an environmental study which contains the ammonium ($NH_4$) concentration (mg/L) in precipitation measured at Olympic National Park, Hoh Ranger Station, weekly or every other week from January 6, 2009 through December 20, 2011. The data set {\tt Olympic.NH4.df} is available in the {\tt EnvStats} R package (Millard, 2013). There are $n = 102$ observations in the data set. Due to the detection limit of the ammonium concentration, 46 of the observations are left-censored.

Let $\delta_{i}$, $i = 1, 2, \ldots, n$ be the left-censored indicator, i.e., $\delta_{i}$ = 1 if the $i$-th observation is left-censored and $\delta_{i} = 0$ otherwise. Then, the likelihood function based on the data $(y_{i}, \delta_{i})$, $i = 1, 2, \ldots, n$ ($n = 102$) can be expressed as
\begin{eqnarray*}
L(\btheta) = \prod_{i=1}^{n} [f(y_{i};\btheta)]^{(1 - \delta_{i})} [F_{Y}(y_{i}; \btheta)]^{\delta_{i}}.
\end{eqnarray*}

We consider two commonly used two-parameter right-skewed distributions with positive support, the Weibull distribution with pdf
$$f(y; \alpha, \beta) = \frac{\alpha}{\beta} \left(\frac{y}{\beta} \right)^{\alpha -1} \exp\left[ \left(\frac{y}{\beta}\right)^{\alpha} \right], y \in \mathbb{R^{+}}, \alpha \in \mathbb{R^{+}}, \beta \in \mathbb{R^{+}}$$
and the lognormal distribution with pdf in Eq. (\ref{lnpdf}). We also consider the TLSN and TLSLG distributions presented in Eq. (\ref{tlsn}) and Eq. (\ref{tlslg}), respectively, for modeling. Once again, we compare the model fitting by using the AIC. The results of the parameter estimates and the values of AIC based on the left-censored data are presented in Table \ref{eg2}.

\begin{table}[h]
\begin{center}
{
\caption{Parameter estimates based on maximum likelihood method and the values of AIC for model fitting of the data of ammonium ($NH_4$) concentration (mg/L) in precipitation using four different skewed distributions }
\label{eg2}
\begin{tabular}{ c  l c } \hline
Distribution &  Parameter estimates                           & AIC    \\ \hline
Weibull      & ${\hat \alpha} = 0.69881$, ${\hat \beta} = 0.01458$                               & 178.5678 \\
Lognormal    & ${\hat \mu}^{*}$ = -4.71449, ${\hat \sigma}^{*}$ = 1.25334                         & 180.3288\\
TLSN         & ${\hat \mu} = 0.06114$, ${\hat \sigma} = 0.05435$, ${\hat \lambda} = 131.23778$ & 153.7112 \\
TLSLG        & ${\hat \mu} = 0.02755$, ${\hat \xi} = 0.02166$, ${\hat \lambda} = 172.78826$      & 160.4412 \\ \hline
\end{tabular}
}
\end{center}
\end{table}
From Table \ref{eg2}, we can observe that the TLSN and TLSLG distributions provide better fit compared to the Weibull and lognormal distributions in terms of the AIC. In between the TLSN and TLSLG distributions, the TLSN distribution gives a slightly better fit compared to the TLSLG distribution.

\section{Concluding Remarks}
\setcounter{equation}{0}
\label{con}

In this article, we study a specific class of univariate and absolutely continuous symmetric skewed probability models by using the technique as described in Ferreira and Steel (2006), namely the TLSS family of distributions.
We discuss some structural properties of the TLSS family including random sample generation from any specific members of the TLSS family of distributions. Interestingly, this class also subsumes some well known skew probability models as particular choices. Since data obtained from environmental studies often follows a skewed distribution, we suggest the use of the TLSS skew symmetric distributions as an alternative probability models to explain random phenomena arising from environmental data. We have used two environmental data sets to illustrate that the TLSS family can be used to model skewed data effectively. The associated inference for such a family of distributions under the Bayesian paradigm will be considered in a future article.


\bibliographystyle{plain}
\bibliography{dissertation}

\end{document}